\documentclass[12pt]{iopart}

\usepackage{graphicx}
\usepackage{bm}

\begin{document}

\title[Multiple radial corrugations in multiwalled carbon nanotubes under pressure]
{Multiple radial corrugations in multiwalled carbon nanotubes under pressure}

\author{Hiroyuki Shima}
\address{Department of Applied Physics, Graduate School of Engineering,
Hokkaido University, Sapporo 060-8628, Japan}
\ead{shima@eng.hokudai.ac.jp}
\author{Motohiro Sato}
\address{Department of Socio-Environmental Engineering, Graduate School of Engineering, Hokkaido University, Sapporo 060-8628 Japan}
\ead{tayu@eng.hokudai.ac.jp}

\begin{abstract}
Radial elastic corrugation
of multi-walled carbon nanotubes under hydrostatic pressure
is demonstrated by using the continuum elastic theory.
Various corrugation patterns are observed
under several GPa,
wherein the stable cross-sectional shape depends on 
the innermost tube diameter $D$
and the total number $N$ of concentric walls.
A phase diagram is established
to obtain the requisite values of $D$ and $N$ for a desired corrugation pattern among choices.
In all corrugation patterns,
the cylindrical symmetry of the innermost tube is maintained even under high external pressures.

\end{abstract}



\submitto{\NT}
\maketitle

\section{Introduction}

Carbon nanotubes show great promise for use as nanoscale materials
due to their high tensile strength in the axial direction 
and remarkable flexibility in bending \cite{Sears2004}.
Another important characteristic of carbon nanotubes
is their high flexibility in the radial direction.
In fact, the magnitude of radial stiffness of an isolated carbon nanotube
is considerably less than that of axial stiffness \cite{Palaci2005},
which allows a reversible change in the cross-sectional shape
on applying a hydrostatic pressure.
Such a pressure-induced radial deformation
and the associated change in vibrational modes
are useful to probe
the structural properties of nanotubes.
More interestingly,
this deformation alters other physical properties of nanotubes and related materials,
such as electronic \cite{Park1999,Mazzoni2000,DS_Tang2000,Gomez2006,Taira2007,Nishio2008,Dandoloff2008}
and optical \cite{Venkateswaran1999,Loa2003,Deacon2006,Lebedkin2006,Onoe2007,Longhurst2007}
properties.
For instance,
application of hydrostatic pressures have been found to induce
drastic changes in the electrical conductance of nanotubes
\cite{Cai2006,Monteverde2006},
implying the relevance of radial deformations
in carbon nanotube applications.

Thus far, many experimental and theoretical studies have been carried out
on radial deformations of carbon nanotubes induced by
hydrostatic pressures
\cite{J_Tang2000,Peters2000,Sharma2001,Rols2001,Reich2002,Elliott2004,Tangney2005,
Gadagkar2006,Wang2003,Zhang2006,Natsuki2006,Hasegawa2006,Yang2006,Chrisofilos2007,
Imtani_2007,Imtani_2008},
most of which have focused on single-walled nanotubes (SWNTs) and their bundles.
Successive investigations have revealed flattening and polygonalization 
in the cross section of SWNTs under pressures of the order a few GPa
\cite{Venkateswaran1999, Peters2000},
in which the stable cross-sectional shape is determined by minimizing
the elastic energy of the cylindrical tube under constraints.
Employing these transformation properties for developing
nanoscale pressure sensors has also been suggested \cite{Wu2004, Mahar2007}.

As compared to the intensive studies carried on SWNTs, 
studies on the radial elastic deformation of
multiwalled nanotubes (MWNTs) are lagging behind.
An important feature of MWNTs is that 
they consist of a set of concentric graphite cylindrical walls
mutually interacting via the van der Waals (vdW) forces.
This core-shell structure produces an encapsulation effect under external pressure,
wherein the outer walls collectively function as a protective shield.
In principle, the encapsulation effect enhances the radial stiffness of MWNTs.
However, this is less obvious if 
the number of concentric walls, $N$, is much greater than unity.
For the latter case,
outside walls possess large diameters so that 
the application of a hydrostatic pressure leads to 
a mechanical instability in the outer walls.
This instability is, nonetheless,
compensated by the relative rigidity of the inner walls with small diameters.
These two competing effects imply
the possibility of a new cross-sectional shape transition
of MWNTs induced by hydrostatic pressure,
whereas such the transition is still to be explored.

\begin{figure}[bbb]
\hspace{0mm}
\vspace{0mm}
\begin{center}
\includegraphics[width=4.6cm]{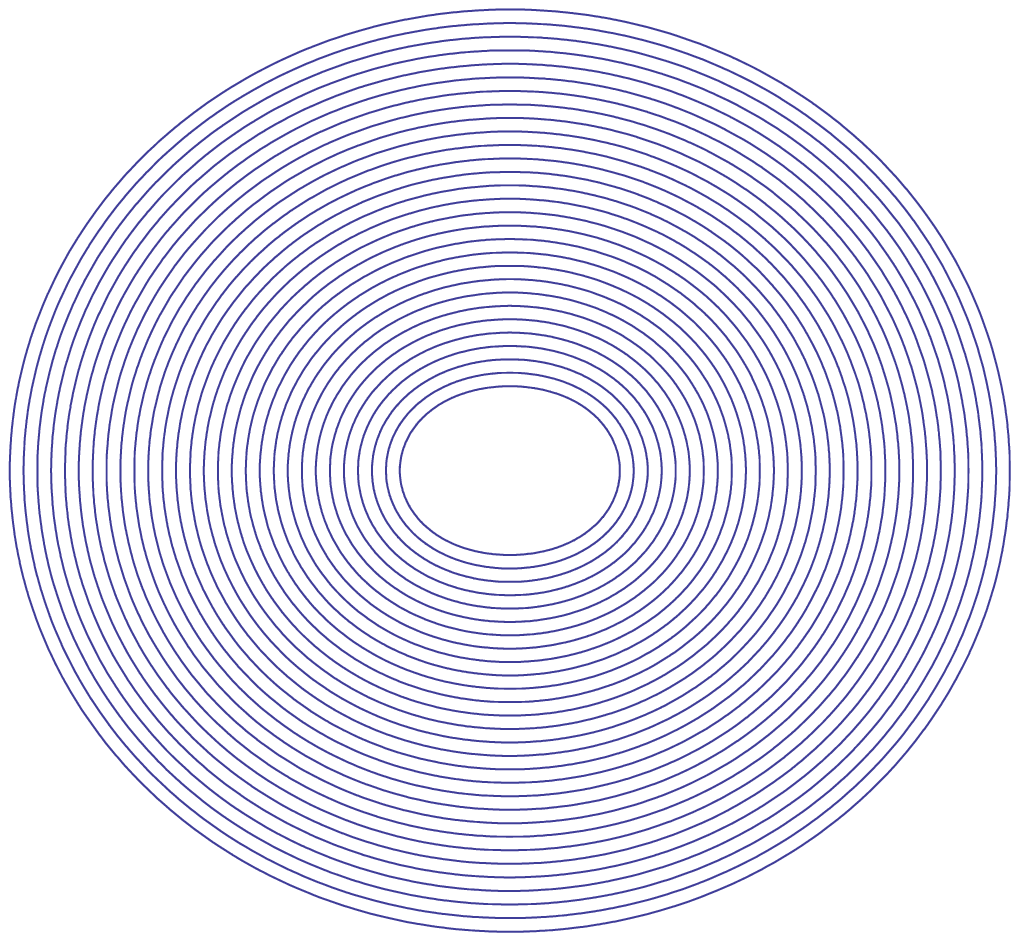}
\includegraphics[width=4.6cm]{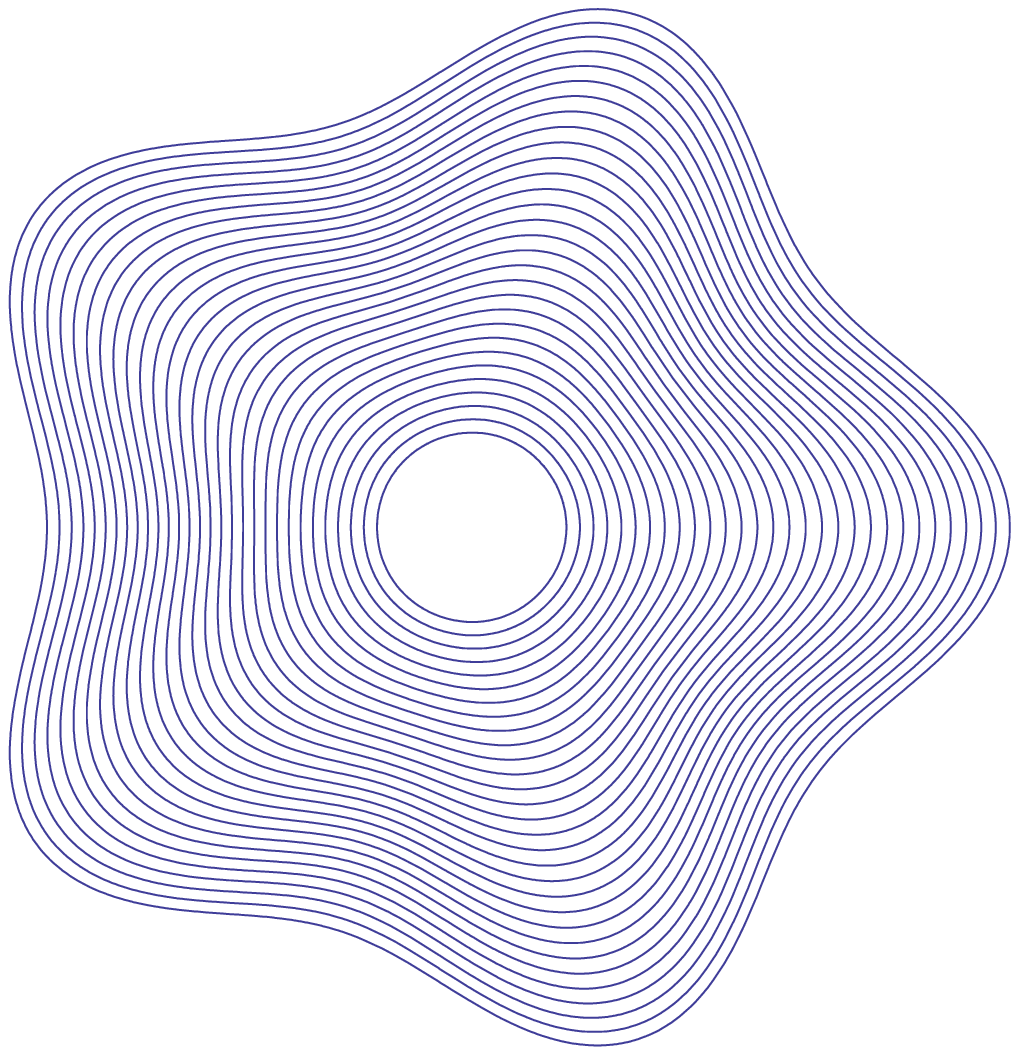}
\end{center}
\vspace{0cm}
\caption{(color online) Cross-sectional views of (a) elliptic $(n=2)$ and
(b) corrugated $(n=5)$ deformation modes observed for
$N$-walled nanotubes with (a) $N=29$ and (b) $N=30$.
The innermost tube diameter $D=5.0$ nm is fixed.
The mode index $n$ indicates the wave number of the deformation mode
along the circumference.}
\label{fig_01}
\end{figure}

In this article, we demonstrate a novel radial deformation,
called the radial corrugation,
of MWNTs with $N\gg 1$ under hydrostatic pressure.
In the corrugation mode,
outside walls show significant radial corrugation
along the circumference,
while the innermost tube maintains its cylindrical symmetry (Fig.~\ref{fig_01}(b)).
We found various corrugation modes can be obtained 
by tuning the innermost tube diameter $D$
and the number of constituent walls $N$,
which is a direct consequence of the core-shell structure of MWNTs.
These results provide useful information for developing
nanofluidic \cite{Majumder2005,Noy2007,Whitby2007,Khosravian2008}
or nanoelectrochemical \cite{Frackowiak2001,Kowalczyk2007} devices
whose performance depends crucially on 
the geometry of the inner hollow cavity of nanotubes.

\begin{figure}[ttt]
\hspace{0mm}
\vspace{0mm}
\begin{center}
\includegraphics[width=8.0cm]{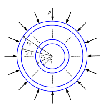}
\includegraphics[width=5.0cm]{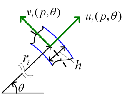}
\end{center}
\vspace{0cm}
\caption{(color online) Illustrations of geometric parameters of our
continuum elastic model.
Left: Sketch of cross section of MWNT subjected to hydrostatic pressure $p$.
Here, $r_i$ indicates the radius of the $i$th cylindrical wall
of the MWNT.
Right: Sketch of displacement of surface element of $i$th wall.
Here, $u(p, \theta)$ and $v(p, \theta)$ denote the deformation amplitudes
of the surface element in the radial and circumferential directions,
respectively.}
\label{fig_02}
\end{figure}

\section{Method}

\subsection{Mechanical energy of MWNT}

The stable cross-sectional shape of a MWNT under a hydrostatic pressure $p$
can be evaluated by using
the continuum elastic theory for cylindrical shells.
The effectiveness of the continuum approximation
for modeling MWNTs has been demonstrated in a series of studies previously
\cite{Ru,Wang,Shen,Rafii,He2005},
particularly for multiwalled nanotubes containing a large number of carbon atoms.
In the continuum approximation,
the mechanical energy $U$ of a MWNT per unit length in the axial direction
is the sum of
the deformation energy $U_D$ of all concentric walls,
the interaction energy $U_I$ of all adjacent pairs of walls,
and the potential energy $\Omega$ of the applied pressure.
All the three energy terms are functions of $p$ and 
the deformation amplitudes $u_i(p,\theta)$ and $v_i(p,\theta)$,
which denote the radial and circumferential displacements, respectively,
of a surface element of the $i$th wall (see Fig.~\ref{fig_02}).
Thus, $U$ can be written as
\begin{equation}
U = U\left[p, u_i(p,\theta), v_i(p,\theta) \right] = U_D + U_I + \Omega.
\label{eq_01}
\end{equation}
It should be noted that 
our theoretical model assumes a very long, straight MWNT 
with both ends being free. On applying hydrostatic pressure, therefore, 
the tube is freely (but slightly) elongated in the longitudinal direction because of 
the absence of reactive force at the ends. 
This means the absence of longitudinal deformation in our theoretical condition, 
which allows us to consider only the cross sectional deformation under hydrostatic pressure.

An explicit form of the function (\ref{eq_01}) is obtained as follows.
Firstly, $U_D$ of all concentric walls
is expressed as
\begin{equation}
U_D = \sum_{i=1}^N
\int_0^{2\pi} 
\left\{
\frac{\alpha_i}{2r_i}
\left[
u_i + {v_i}' + \frac{\left({u_i}' - v_i\right)^2}{2 r_i}
\right]^2 
+
\frac{\beta_i}{2}
\frac{ \left( {u_i}'' - {v_i}' \right)^2 }{r_i^3}
\right\}
d\theta,
\label{eq_01aa}
\end{equation}
where $u' \equiv du/d\theta$.
In Eq.~(\ref{eq_01aa}),
the first term in the curly brackets
describes the stretching energy of the walls
in the tangential direction
and the second term describes the bending energy.
(The derivation of the formula (\ref{eq_01aa}) involves some complex calculations
and is presented in Appendix.)
The two parameters $\alpha_i$ and $\beta_i$ characterise the mechanical
stiffness of the walls for in-plane stretching and bending, respectively.
They are defined as follows:
$$
\alpha_i = \frac{Eh}{1-\nu^2}
\;\;\; \mbox{and} \;\;\;
\beta_i = \frac{E h^3}{12\left(1-\nu^2 \right)},
$$
where $E$ is Young's modulus ($E = 1$ TPa),
$\nu$ is Poisson's ratio ($\nu= 0.27$),
and $h$ indicates the thickness of individual walls ($h = 0.34$ nm).
Hereafter, $r_i$ indicates the radius of the $i$th wall 
in the absence of pressure; thus $D = 2 r_1$.
The spacing between adjacent walls is set to be
$|r_i - r_{i\pm 1}| = 0.344 + 0.1 {\rm e}^{-D/2}$ nm
according to Ref.~\cite{Kiang}.

Secondly, the explicit form of $U_I$ is given as
\begin{equation}
U_I = \sum_{i=1}^{N-1} 
\frac{c_{i,i+1} r_i}{2}
\int_0^{2\pi} 
\left( u_i - u_{i+1} \right)^2 d\theta
+
\sum_{i=2}^N
\frac{c_{i,i-1} r_i}{2}
\int_0^{2\pi} 
\left( u_i - u_{i-1} \right)^2 d\theta,
\end{equation}
where the vdW interaction coefficients $c_{i,j}$ are functions of $r_i$ and $r_j$
and given as follows \cite{He2005}:
$$
c_{ij} = 
- \left( 
\frac{1001 \pi \varepsilon \sigma^{12}}{3 a^4} F_{ij}^{13}
- \frac{1120 \pi \varepsilon \sigma^6}{9 a^4} F_{ij}^7
\right) r_j.
$$
Here we set
$$
F_{ij}^m = \frac{1}{\left( r_i + r_j \right)^m} \int_0^{\pi/2} 
\frac{d\theta}{\left( 1- K_{ij} \cos^2 \theta\right)^{m/2}}
\quad {\rm and} \quad
K_{ij} = \frac{4 r_i r_j}{\left( r_i + r_j \right)^2},
$$
in which $a$ denotes the chemical bond length between neignbouring carbon atoms 
within a layer ($a = 0.142$ nm),
and $\varepsilon$ and $\sigma$ are the parameters
that determine the vdW interaction between two layers 
($\varepsilon = 2.968$ meV and $\sigma = 0.3407$ nm) \cite{Saito}.

Finally, we consider an explicit form of $\Omega$.
Since a cylindrical shell subjected to $p$
is a conservative system,
the change in the potential energy as the cross section of the MWNT deforms
is the negative of the work done by the pressure during the deformation.
Hence, we have
\begin{equation}
\Omega = - p(\pi r_N^2 - S^*),
\end{equation}
where $S^*$ is the cross-sectional area after deformation.
A simple calculation yields the following expression:
\begin{equation}
\Omega = p \int_0^{2\pi} 
\left( r_N u_N + \frac{u_N^2 + v_N^2 - {u_N}' v_N + u_N {v_N}'}{2} \right) d\theta.
\label{eq_omega}
\end{equation}
Refer Appendix for the derivation of the formula (\ref{eq_omega}).

\subsection{Critical pressure $p_c$ and deformation mode $n$}

Our aim is to evaluate the critical pressure $p_c$
above which the circular cross section of MWNTs is elastically deformed
into non-circular one.
To carry out the analysis,
we decompose the radial displacement terms as 
$u_i(p,\theta) = u_i^{(0)} (p) + \delta u_i(\theta)$.
Here, $u_i^{(0)} (p)$ indicates a uniform radial contraction
of the $i$th wall at $p<p_c$,
whose magnitude is proportional to $p$.
The $\delta u_i(\theta)$ describes a deformed, non-circular cross section
observed just above $p_c$.
Similarly, we can write
$v_i(p,\theta) = \delta v_i(\theta)$, since $v_i^{(0)}(p)\equiv 0$ at $p<p_c$.

Applying the variation method to $U$ with respect to $u_i$ and $v_i$,
we obtain a system of $2N$ linear differential equations
given by
\begin{eqnarray}
& &
\alpha_i ( \delta u_i + \delta {v_i}' - \gamma_i \eta_i' )
+ \beta_i \eta_i''' 
+
p \delta_{i,N} (\delta u_i + \delta v_i') \nonumber \\
& &+
\left( 1-\delta_{i,N} \right) c_{i,i+1} r_i 
\left( \delta u_i - \delta u_{i+1} \right) \nonumber \\
& &+
\left( 1-\delta_{i,1} \right) c_{i,i-1} r_i 
\left( \delta u_i - \delta u_{i-1} \right) = 0, \quad (i=1,\cdots, N)
\label{eq_02}
\end{eqnarray}
and
\begin{equation}
\alpha_i 
\left(
\delta {u_i}' + \delta {v_i}'' + \gamma_i \eta_i
\right) - \beta_i \eta_i'' 
+ p \delta_{i,N} (\delta {u_i}' - \delta v_i) = 0,  \quad (i=1,\cdots, N)
\label{eq_02xx}
\end{equation}
where $\gamma_i = u_i^{(0)}(p)/r_i$ and $\eta_i = u_i' -v_i$.
In deriving Eqs.~(\ref{eq_02}) and (\ref{eq_02xx}) through the calculus of variation,
quadratic or cubic terms in $\delta u_i$ and $\delta v_i$ were omitted
since we consider elastic deformation with sufficiently small displacements.
In addition, the terms consisting only of $u_i^{(0)}$ and $p$ are also omitted;
the sum of such terms should be equal to zero since $u_i^{(0)}$ represents
an equilibrium circular cross-section under $p$.\footnote{The fact that 
the sum equals zero determines the function form of $u_i^{(0)}(p)$.}

Clearly, $\delta u_i$, $\delta v_i$ and their derivatives 
are periodic in $\theta$.
Hence, general solutions of Eqs.~(\ref{eq_02}) and (\ref{eq_02xx}) are given by
the Fourier series expansions
$$
\delta u_i(\theta) = \sum_{n = 1}^{\infty} \delta \bar{\mu}_i(n) \cos n\theta
\;\;\; \mbox{and} \;\;\;
\delta v_i(\theta) = \sum_{n = 1}^{\infty} \delta \bar{\nu}_i(n) \sin n\theta.
$$
Introduction into Eqs.~(\ref{eq_02}) and (\ref{eq_02xx}) results in
the matrix equation $\bm{C} \bm{u} = \bm{0}$,
in which the vector $\bm{u}$ consists of
$\delta \bar{\mu}_i(n)$ and $\delta \bar{\nu}_i(n)$ with all possible $i$ and $n$,
and the matrix $\bm{C}$ involves one variable $p$
as well as parameters such as $\alpha_i$, $\beta_i$, $\cdots$ {\it etc}.
It should be noted that,
due to the orthogonality of $\cos n\theta$ and $\sin n\theta$,
the matrix $\bm{C}$ can be expressed by a block diagonal matrix 
of the form $\bm{C} = \bm{C}_{n=1} \oplus \bm{C}_{n=2} \oplus \cdots$.
Here, $\bm{C}_{n=m}$ is a $2N \times 2N$ submatrix
that satisfies $\bm{C}_{n=m} \bm{u}_{n=m} = \bm{0}$,
where $\bm{u}_{n=m}$ is a $2N$-column vector composed of
$\delta \bar{\mu}_i(n=m)$ and $\delta \bar{\nu}_i(n=m)$.
As a result,
the secular equation ${\rm det}(\bm{C}) = 0$ that provides 
nontrivial solutions of Eqs.~(\ref{eq_02}) and (\ref{eq_02xx}) is rewritten by
\begin{equation}
{\rm det} \left( \bm{C}_{n=1} \right) {\rm det} \left( \bm{C}_{n=2} \right) \cdots = 0.
\label{eq_det}
\end{equation}
Solving Eq.~(\ref{eq_det}) with respect to $p$,
we obtain a sequence of discrete values of $p$
each of which is the smallest solution of ${\rm det} \left( \bm{C}_{n=m} \right) \cdots = 0$
$(m=1,2,\cdots)$.
Among these values of $p$s, the minimum one serves as the critical pressure $p_c$
that is associated with a specific integer $n=m$.
From the definition, the $p_c$ associated with a specific $m$
allows only $\delta \bar{\mu}_i(n=m)$ and $\delta \bar{\nu}_i(n=m)$ be finite,
but it requires $\delta \bar{\mu}_i(n\ne m) \equiv 0$ and 
$\delta \bar{\nu}_i(n\ne m) \equiv 0$.
Immediately above $p_c$, therefore, the circular cross section of MWNTs
becomes radially deformed as described by
$$
u_i(\theta) = u_i^{(0)}(p_c) + \delta \bar{\mu}_i(n) \cos n\theta
\;\;\; \mbox{and} \;\;\;
v_i(\theta) = \delta \bar{\nu}_i(n) \sin n\theta,
$$
where the value of $n$ is uniquely determined by 
the one-to-one relation between $n$ and $p_c$.

\section{Radial corrugations of MWNTs}

\subsection{Cross-sectional view}

Figure \ref{fig_01} illustrates cross-sectional view of two typical deformation modes ---
(a) elliptic $(n=2)$ and (b) corrugation $(n=5)$ modes --- 
of a MWNT with $D=5.0$ nm.
For Fig.~\ref{fig_01}(a) and 1(b), $N=29$ and $30$, respectively.
In the elliptic mode, all constituent walls are radially deformed.
On the contrary, in the corrugation mode,
outside walls exhibit significant deformation,
while the innermost wall maintains its circular shape.
The deformation mode observed just above $p_c$
depends on the values of $N$ and $D$
for the MWNT under consideration.
It will be shown below that larger $N$ and smaller $D$ favor
a corrugation mode with larger $n$.

\begin{figure}[ttt]
\hspace{-5mm}
\vspace{0cm}
\begin{center}
\includegraphics[width=9cm]{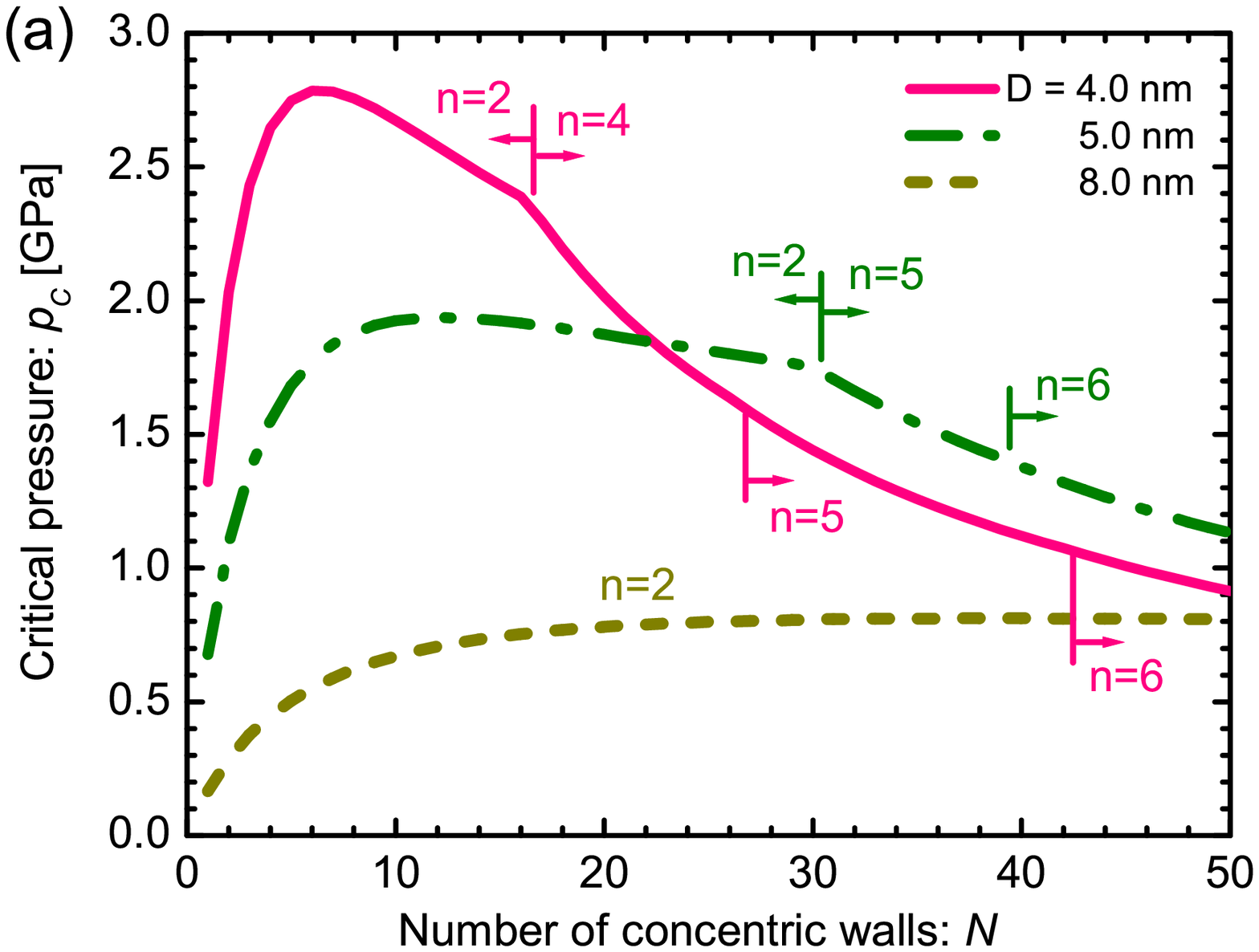}
\includegraphics[width=9cm]{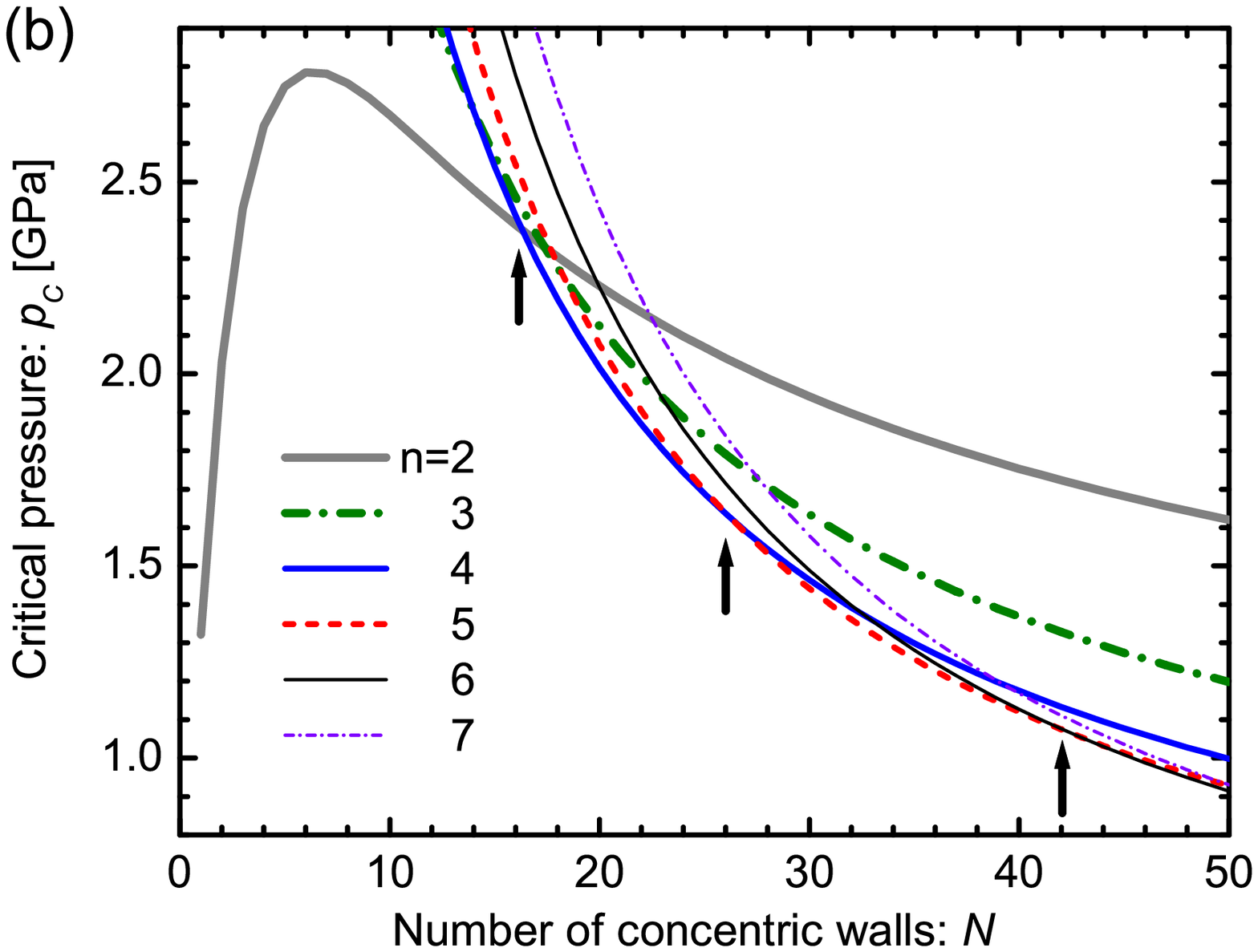}
\end{center}
\vspace{0cm}
\caption{(color online) (a) Critical pressure curves showing
$p_c$ required to produce radial deformation
of $N$-walled nanotubes with fixed $D$.
The mode index $n$ that characterizes the radial deformation mode
observed just above $p_c$ is also shown.
(b) Branches of solutions $p(N)$ for 
secular equation ${\rm det} (\bm{C}) = 0$ (refer text).
The $N$-dependence of $p(N)$ for each deformation mode $n$ is displayed.
The innermost tube diameter is set to be $D=4.0$ nm for all curves.
For a fixed $N$, the minimum value of $p$ among the branches
functions as the critical pressure $p_c(N)$
just above which radial deformation takes place.
The vertical arrows indicate the phase boundary points 
across which the deformation modes change.
}
\label{fig_03}
\end{figure}

\subsection{Critical pressure curve}

Figure \ref{fig_03} (a) shows $p_c$ as a function of $N$ for various values of $D$.
The mode index $n$ of the deformation mode
observed just above $p_c$ for fixed $N$ and $D$ is also shown.
For all $D$, $p_c$ increase with $N$ followed by a slow decay,
except for the case of $D=8.0$ nm.
The increase in $p_c$ in the region of small $N$ is attributed to 
the enhancement of radial stiffness of the entire MWNT by encapsulation.
This stiffening effect disappears with further increase in $N$,
resulting in the decay of $p_c(N)$.
A decay in $p_c$ implies that a relatively low pressure becomes sufficient
to produce radial deformation, thus indicating
an effective ``softening" of the MWNT.
Such a decay is also observed for $D=8.0$ nm and larger $D$, in principle,
if a sufficiently large $N$ is considered (but omitted in Fig.~\ref{fig_03} (a)).
The two contrasting behaviors, stiffening and softening,
are different manifestations of the encapsulation effect of MWNTs.

We observe in Fig.~\ref{fig_03} (a), only the elliptic deformation is realised
in the stiffened region,
while certain radial corrugations with $n\ge 3$ are formed in the softened region.
The observation of the former phenomenon can be understood 
by considering the radial deformation of a 
single-walled cylindrical shell.
In the stiffened region, since $N$ is very small,
the relation $|r_1-r_N|\ll r_N$ holds.
This implies that in this region, a MWNT
behaves effectively as a single-walled cylindrical shell
with wall thickness $|r_1-r_N|$.
For a single-walled shell,
the secular equation ${\rm det} (\bm{C})= 0$ gives
solutions of
$$
p = \frac{E h^3}{12 \left( 1-\nu^2 \right) r_N^3} 
\; \frac{n^2-1}{1+ (h^2/12 r_N^2)} \quad (n=2,3,\cdots).
$$
Hence, $p$ always has the minimum value when $n=2$.
This scenario applies to MWNTs if $N$ is sufficiently small,
resulting in the elliptic mode of $n=2$ in the stiffened region.

The most striking observation is the
successive transformation of the cross section with an increase in $N$.
For $D=4.0$ nm, for example,
the deformation mode observed just above $p_c$ jumps abruptly 
from $n=2$ to $4$ at $N=17$,
followed by successive emergences of higher corrugation modes with larger $n$.
The critical number of walls $N=N_c$ separating the elliptic phase
from the corrugation phase
is identified to $N$ that yields a cusp in the curve of $p_c(N)$.
In contrast, no singularity is observed in the curve of $p_c(N)$
at values of $N$, which separate two neighboring corrugation phases.
We emphasize that at these phase boundaries,
one additional wall induces a drastic change in 
the cross-sectional shape of the MWNT under the hydrostatic pressure.
These transitions in $n$ originate from the two competing effects
inherent in MWNTs with $N\gg 1$, that is,
the relative rigidity of the inner walls
and the mechanical instability of outer walls.
A large discrepancy in the radial stiffness of the inner and outer walls
gives rise to a maldistribution of the deformation amplitudes
of concentric walls interacting through the vdW forces,
which consequently produces an abrupt change in the observed deformation mode at some $N$.

Figure \ref{fig_03}(b) explains the absence of the corrugation mode of $n=3$
in MWNTs with $D=4.0$ nm.
This figure shows the $N$-dependence of the solutions $p(N)$ 
for the secular equation
${\rm det}(\bm{C}) = 0$.
As mentioned earlier, the secular equation gives various values of $p$,
each of which is associated with a specific mode index $n$.
Among the values of $p$,
the minimum value gives the critical pressure $p_c$ just above which
cross-sectional deformation takes place.
Figure \ref{fig_03}(b) depicts the $N$-dependence of $p(N)$ for several $n$ values,
where the innermost tube radius is fixed to be $D=4.0$ nm.
For $N< 17$,
the values of $p$ for $n=2$ are less than those for $n\ge 3$,
which implies that the elliptic mode occurs for MWNTs with $N<17$.
However, for $N\ge 17$, the minimum $p$ corresponds to $n=4$,
implying the occurrence of the corrugation mode of $n=4$.
It should be noted that for $n=3$, $p$ can never attain the minimum value
at any $N$.
This is why the corrugation mode of $n=3$ cannot be observed
for MWNTs with $D=4.0$ nm.
A parallel discussion accounts for the absence of the modes of $n=3,4$ for $D=5.0$ nm
and the corrugation modes for $D=8.0$ nm within the range of $N$ we have considered.
It also follows from Fig.~\ref{fig_03}(b) that 
the cusps in the curves $p_c(N)$ occur only at the phase boundary $N_c$
separating the elliptic phase $(n=2)$ from a corrugation phase $(n>3)$,
while no singularity appears at the boundaries of $N$ between neighbouring corrugation phases.

\begin{figure}[ttt]
\hspace{-5mm}
\vspace{0cm}
\begin{center}
\includegraphics[width=8.0cm]{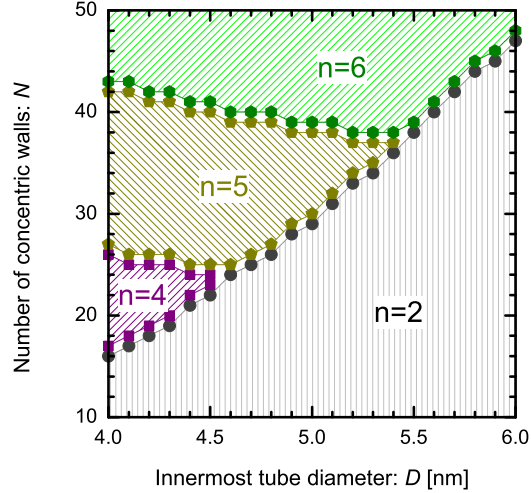}
\end{center}
\vspace{0cm}
\caption{(color online) Phase diagram of radial deformation modes
observed above $p_c$.
Various corrugation modes are obtained depending on the values of $N$ and $D$.}
\label{fig_04}
\end{figure}

\subsection{Phase diagram}

Figure \ref{fig_04} shows a phase diagram of the radial deformation modes in MWNTs
observed above $p_c$.
The bottom region below a chain of solid circles (colored in gray online)
corresponds to the elliptic phase,
and the top regions surrounded by other symbols indicate the corrugation phases
associated with $n$.
It is clearly observed that above the elliptic phase,
multiple corrugation modes are formed depending on the values of $N$ and $D$.
It is also observed that smaller $D$ and larger $N$
favor corrugation modes to the elliptic modes.
Furthermore, larger $N$ yield higher corrugation modes with larger $n$.
To our knowledge,
this is the first demonstration of the phase diagram of the multiple corrugation transitions,
while conditions for the higher modes to occur
were previously argued \cite{Wang2003}.
A quantitative examination of the present results
by sophisticated atomistic simulations \cite{XYLi, Arias, Arroyo}
may provide a better understanding of the corrugation properties of MWNTs.

\section{Persistent cylindrical geometry of the innermost tube}

\begin{figure}[ttt]
\vspace{0cm}
\begin{center}
\includegraphics[width=10cm]{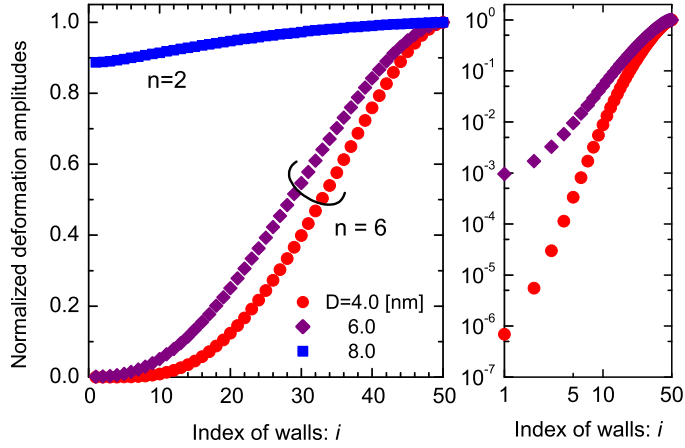}
\end{center}
\vspace{0cm}
\caption{(color online)
Left: Deformation amplitude of each $i$th concentric wall of MWNT with $N=50$.
The mode index $n$ of the associated deformation mode is also indicated.
Right: Double-logarithmic plots of data plotted in left-hand-side figure.}
\label{fig_05}
\end{figure}

We have mentioned that in all corrugation modes,
the innermost tube maintains its cylindrical symmetry.
This persistence of cylindrical symmetry can be
demonstrated by plotting the deformation amplitudes
$\delta \bar{\mu}_i$.
Figure \ref{fig_05} shows
the normalized deformation amplitudes, $|\delta \bar{\mu}_i/\delta \bar{\mu}_N|$,
of individual concentric walls for the MWNT with $N=50$.
The innermost and outermost walls correspond to $i=1$ and $i=50$, respectively,
and the associated $n$ is also shown.
Double-logarithmic plots of the same data for the corrugation mode $(n=6)$
are also presented in the right-hand-side figure.
The figures clearly show that in the corrugation mode,
the deformation amplitudes of the innermost tube are significantly
smaller than those of the outer walls by several orders of magnitude.
The persistence of cylindrical symmetry of the innermost tube
will be useful in developing nanotube-based nanofluidic
\cite{Majumder2005,Noy2007,Whitby2007,Khosravian2008}
or nanoelectrochemical devices \cite{Frackowiak2001,Kowalczyk2007},
since both utilize the hollow cavity within the innermost tube.
In fact, several different types of intercalated molecules
such as diatomic gas, water, organic, and transition metal molecules
are known to fill the innermost hollow cavities of nanotubes \cite{Noy2007}
and exhibit various intriguing behaviors
that are distinct from those of 
the corresponding bulk systems \cite{Monthioux2002, CKYang2003, Joseph2008}.
For the intercalates,
the innermost tube of MWNTs with $N\gg 1$
functions as an ideal protective shield,
since it maintains its cylindrical geometry 
even under high external pressures up to an order of GPa.

\section{Discussions}

We have observed that radial corrugation modes yield 
a large difference in the deformation amplitudes between the outer and inner walls.
Hence, the modes disturb the equal spacings between the concentric walls in MWNTs.
This possibly affects electronic and vibrational properties of the entire nanotube,
thus triggering a change in its electronic and thermal conductance.
Such pressure-induced changes, if they occur,
are of practical use for developing MWNT-based pressure sensors \cite{Wu2004, Mahar2007}.
Besides, these changes should be relevant to
the characteristics of carbon nanotube composites \cite{Coleman2006,Li2008}
that contain many clusters of MWNTs dispersed in an elastic medium.
In the composites,
the vdW interaction between adjacent tubes
is sufficiently strong to yield radial corrugation,
which affects the mechanical strength or the conducting properties of the composites.
Intensive studies on these issues are expected to yield
new methods for engineering next-generation devices and nanomaterials.

Before presenting the conclusion, 
we mention the relevance of the lattice registry 
in radial corrugation phenomena.
The atomic lattice registry 
({\it i.e.}, the degree of commensurance in atomic structures
between neighbouring carbon layers)
is known to play a prominent role
in determining the optimal morphology of fully collapsed MWNTs
\cite{Yu_regist,Liu_regist,Xiao_regist}.
For a similar reason,
it possibly affects the cross-sectional shape of the corrugation modes,
if the interlayer spacings partially vanish by applying pressure
much higher than that we have considered;
Crumpling or twisting in outside walls may be observed
depending on the atomic configuration of the MWNT.
We also conjecture that the difference in the lattice registry
between adjacent layers will induce a shift in the phase boundary
depicted in the phase diagram of Fig.~\ref{fig_04}. 
For instance, the AA stacking of carbon atoms \cite{Xiao_regist}
in the radial direction will lead to an effective load transfer 
between neighbouring layers.
Thereby, rigid inner walls can support effectively
the mechanical instability of outer walls. 
This implies that the phase boundary curve shown in Fig.~\ref{fig_04}
separating the elliptic phase ($n=2$) from corrugation 
phases ($n\ge 3$) shifts downward.
On the other hand, the AB stacking tends to shift upward 
toward the phase boundary since the rigid inner walls provide
minimal support.
The magnitude of the phase boundary shift will be 
quantitatively determined in our future studies.

\section{Conclusion}

In conclusion, we have demonstrated the presence of multiple radial corrugations
peculiar to MWNTs under hydrostatic pressures.
Theoretical investigations based on the continuum elastic theory have revealed that
MWNTs consisting of a large number
of concentric walls undergo elastic deformations at critical pressure $p_c \sim$ 1 GPa,
above which the cross-sectional circular shape becomes radially corrugated.
A phase diagram has been established
to obtain the requisite values of $N$ and $D$ for observing a desired corrugation mode.
It is remarkable that in all corrugation modes,
the cylindrical symmetry of the innermost tube is maintained
even under high external pressures.
This persistence of the cylindrical symmetry 
of the innermost tube of MWNTs is completely 
in contrast to the pressure-induced collapse of SWNTs.
We believe that a study of this behavior of MWNTs 
will shed light on the potential of MWNT-based devices.

\section*{Acknowledgments}

This study was supported by a Grant-in-Aid for Scientific
Research from the MEXT, Japan.
One of the authors (H.S.) is thankful for the financial support from the
Sumitomo Foundation.
A part of the numerical simulations were carried using
the facilities of the Supercomputer Center, ISSP, University of Tokyo.

\begin{figure}[ttt]
\vspace{0cm}
\begin{center}
\includegraphics[width=5.5cm]{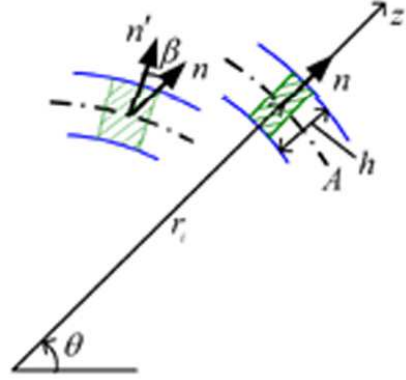}
\end{center}
\vspace{0cm}
\caption{(color online) Part of cross section of $i$th cylindrical wall 
undergoing radial deformation.
Cross-sectional deformation leads to
the translation of the shaded region colored in dark green
(located at the point $(r_i, \theta)$) into the adjacent shaded region
colored in light green.
The vectors $\bm{n}$ and $\bm{n}'$ indicate the normals to the centroidal surface
(dashed-dotted) before and after deformation, respectively.
The shaded region after deformation exhibits stretching in the circumferential direction 
with rotation by $\beta$.
}
\label{fig_app}
\end{figure}

\section*{Appendix}
\appendix
\setcounter{section}{1}

\subsection{Derivation of the deformation energy $U_D$}

In this section, we provide the explicit derivation of $U_D$
introduced in Eq.~(\ref{eq_01aa}).
Consider the $i$th cylindrical wall of a  long and thin circular tube with thickness $h$.
A surface element of the cross-sectional area of the wall is expressed by $(r_i d\theta)dz$,
where $\theta$ is the circumferential angle around the cylindrical axis
and $z$ is a radial coordinate measured from the centroidal surface
(indicated by $A$ in Fig.~\ref{fig_app}).
The stiffness of the element for stretching along the circumferential direction
is given by $E/(1-\nu^2)$, where $E$ and $\nu$ are
Young's modulus and Poisson's ratio, respectively, of the wall.
Thus, the deformation energy $U_D^{(i)}$ of the $i$th wall per unit length
in the axial direction
is written as
\begin{equation}
U_D^{(i)} = \frac{E r_i}{2 \left( 1-\nu^2 \right)} 
\int_{-h/2}^{h/2} \int_0^{2\pi} \bar{\varepsilon}(z,\theta)^2 dz d\theta.
\label{eq_A2_004}
\end{equation}
Here, $\bar{\varepsilon}(z,\theta)$ is the extensional strain of the element
in the circumferential direction.
For the calculation,
we adopt the strain-displacement relation introduced by Sanders \cite{Sanders}
subject to the the following two assumptions: (i) strains are small and
(ii) the rotation angles $\beta$ of the normal to the centroidal surface
are `small but finite'.
Then, we obtain the relation \cite{Sanders}
\begin{equation}
\bar{\varepsilon}(z, \theta) = \varepsilon(\theta) + z \kappa(\theta),
\label{eq_A2_005}
\end{equation}
where $\varepsilon$ and $\kappa$ are expressed in terms of 
$u_i(\theta)$ and $v_i(\theta)$ as follows:
\begin{equation}
\varepsilon = \frac{u_i + {v_i}'}{r_i} 
+
\frac12 \left( \frac{u_i' - v_i}{r_i} \right)^2
\;\;\; \mbox{and} \;\;\
\kappa = - \frac{{u_i}'' - {v_i}'}{r_i^2}.
\label{eq_A2_006}
\end{equation}
The relation (\ref{eq_A2_005}) states that
the circumferential strain $\bar{\varepsilon}$ of a volume element
can be decomposed into in-plane stretching $\varepsilon(\theta)$
and bending-induced stretching $z \kappa(\theta)$.
The presence of the latter is implicitly illustrated in Fig.~\ref{fig_app},
where the outermost surface is more elongated in the circumferential direction
than the innermost surface due to the thickness of the wall.

From (\ref{eq_A2_004}) and (\ref{eq_A2_005}), we obtain
\begin{equation}
U_D^{(i)} = 
\frac{Eh r_i}{2 \left( 1-\nu^2 \right)} \int_0^{2\pi} \varepsilon^2 d\theta
+
\frac{Eh^3 r_i}{24 \left( 1-\nu^2 \right)} \int_0^{2\pi} \kappa^2 d\theta.
\label{eq_A2_007}
\end{equation}
By substituting (\ref{eq_A2_006}) into (\ref{eq_A2_007}),
followed by some calculations,
we obtain the desired form of $U_D = \sum_{i=1}^N U_D^{(i)}$
as given in (\ref{eq_01aa}).

\subsection{Derivation of the potential energy $\Omega$}

This section presents the derivation of 
the potential energy $\Omega$ of applied pressure
given in Eq.~(\ref{eq_omega}).
Suppose that an elastic cylindrical shell having a radius $r_N$
is subjected to a hydrostatic pressure $p$.
Since $\Omega$ is the negative of the work done by the external pressure
during cross-sectional deformation,
it is expressed as
\begin{equation}
\Omega = - p (\pi r_N^2 - S^*),
\label{eq_app_0}
\end{equation}
where $S^*$ is the cross sectional area of the shell after the deformation
(the sign of $p$ is assumed to be positive inward).

We now consider the explicit functional form of $S^*$.
We denote the Cartesian coordinates of a point 
on the circumference of the cross section by $(x^*, y^*)$.
After the deformation,
they are parametrised by the angle $\theta$ as follows:
\begin{eqnarray}
x^*(\theta) &=& 
\left[ r_N + u_N(\theta) \right] \cos \theta - v_N(\theta) \sin \theta, \nonumber \\
y^*(\theta) &=& 
\left[ r_N + u_N(\theta) \right] \sin \theta + v_N(\theta) \cos \theta.
\label{eq_app_1}
\end{eqnarray}
Here, $u_N(\theta)$ and $v_N(\theta)$ are the deformation amplitudes of a surface element
of the shell in the radial and circumference directions, respectively.
$S^*$ is obtained by the line integral 
around the circumference $C$ as follows:
\begin{equation*}
S^* = \frac12 \oint_C \left( -y^* d x^* + x^* dy^* \right),
\end{equation*}
or equivalently from,
\begin{equation}
S^* = \frac12 \int_0^{2\pi} 
\left( -y^* \frac{d x^*}{d\theta} + x^* \frac{dy^*}{d\theta} \right)
d \theta.
\label{eq_app_2}
\end{equation}
Substituting (\ref{eq_app_1}) into (\ref{eq_app_2}) yields
\begin{equation}
S^* = \pi r_N^2 + \frac12 \int_0^{2\pi} 
\left(
2 r_N u_N + u_N^2 + v_N^2 
- \frac{d u_N}{d\theta} v_N + u_N \frac{d v_N}{d\theta}
\right) d\theta,
\label{eq_app_3}
\end{equation}
where the periodicity relation 
$$
\int_0^{2\pi} \frac{d v_N}{d\theta} d\theta = 0
$$
is employed.
From (\ref{eq_app_0}) and (\ref{eq_app_3}), we finally obtain the desired result
\begin{equation}
\Omega = p \int_0^{2\pi} 
\left( r_N u_N + \frac{u_N^2 + v_N^2 - {u_N}' v_N + u_N {v_N}'}{2} \right) d\theta,
\end{equation}
where $u' \equiv d u/d\theta$.

\end{document}